\documentclass[sigconf,natbib=true]{acmart}
\usepackage{graphicx} 
\usepackage[breakable]{tcolorbox}
\usepackage{algorithm}
\usepackage{algpseudocode}
\usepackage{multirow}
\usepackage{booktabs}
\usepackage{wrapfig}
\usepackage{subfig}

\usepackage{makecell}
\AtBeginDocument{%
  }

\setcopyright{acmlicensed}
\copyrightyear{2018}
\acmYear{2018}
\acmDOI{XXXXXXX.XXXXXXX}
\acmConference[SIGIR '25]{}{July 13--18,
  2025}{Padua, Italy}
\acmISBN{978-1-4503-XXXX-X/2018/06}

\begin{document}

\title{SAGraph: A Large-Scale Social Graph Dataset  with Comprehensive Context for Influencer Selection in Marketing}

\author{Xiaoqing Zhang}
\affiliation{%
  \institution{Gaoling School of Artificial Intelligence, Renmin University of China}
  \city{Beijing}
  \country{China}
}
\email{xiaoqingz@ruc.edu.cn}

\author{Yuhan Liu}
\affiliation{%
  \institution{Gaoling School of Artificial Intelligence, Renmin University of China}
  \city{Beijing}
  \country{China}
}
\email{yuhan.liu@ruc.edu.cn}

\author{Jianzhou Wang}
\affiliation{%
  \institution{Moonshot AI}
  \city{Beijing}
  \country{China}
}
\email{wangjianzhou@msh.team}

\author{Zhenxing Hu}
\affiliation{%
  \institution{Moonshot AI}
  \city{Beijing}
  \country{China}
}
\email{huzhenxing@msh.team}

\author{Xiuying Chen}
\authornote{Corresponding authors.}
\affiliation{%
  \institution{Mohamed bin Zayed University of Artificial Intelligence}
  \city{Abu Dhabi}
  \country{UAE}
}
\email{xiuying.chen@mbzuai.ac.ae}

\author{Rui Yan}
\affiliation{%
  \institution{ Gaoling School of Artificial Intelligence, Renmin University of China}
  \city{Beijing}
  \country{China}
}
\email{ruiyan@ruc.edu.cn}
\authornotemark[1]








\renewcommand{\shortauthors}{Zhang et al.}

\begin{abstract}
Influencer marketing campaign success heavily depends on identifying key opinion leaders who can effectively leverage their credibility and reach to promote products or services. The selecting influencers process is vital for boosting brand visibility, fostering consumer trust, and driving sales. While traditional research often simplifies complex factors like user attitudes, interaction frequency, and advertising content, into simple numerical values. However, this reductionist approach fails to capture the dynamic nature of influencer marketing effectiveness.
To bridge this gap, we present SAGraph, a novel comprehensive dataset from Weibo that captures multi-dimensional marketing campaign data across six product domains. The dataset encompasses 345,039 user profiles with their complete interaction histories, including 1.3M comments and 554K reposts across 44K posts, providing unprecedented granularity in influencer marketing dynamics. SAGraph uniquely integrates user profiles, content features, and temporal interaction patterns, enabling in-depth analysis of influencer marketing mechanisms. Experimental results using both traditional baselines and state-of-the-art large language models (LLMs) demonstrate the crucial role of content analysis in predicting advertising effectiveness. Our findings reveal that LLM-based approaches achieve superior performance in understanding and predicting campaign success, opening new avenues for data-driven influencer marketing strategies.
We hope that this dataset will inspire further research: \url{https://github.com/xiaoqzhwhu/SAGraph/}.
\end{abstract}



\maketitle

\begin{table}[t]
    \centering
    \caption{Comparison of related datasets. `Net' denotes the graph structure. `Interaction' denotes the interaction between different nodes in the Net. `Text' denotes the interaction details that are described in text format.}
    \small
\resizebox{\linewidth}{!}{
\begin{tabular}{cccccc}
\toprule
\textbf{Dataset} & \textbf{Scenario} & \textbf{Net} & \textbf{Interaction} & \textbf{Text} &\textbf{Nodes}\\
\midrule
Blog~\cite{leskovec2007cost}  & knowledge & \checkmark  & $\times$ & $\times$ & 45,000 \\
Reddit~\cite{dutta2020deep} & knowledge & $\times$ & \checkmark & \checkmark & - \\
Wiki Vote~\cite{leskovec2014snap} &rumor& \checkmark  & \checkmark & $\times$&7,115\\
Tweet List~\cite{leskovec2014snap} &rumor& \checkmark & $\times$ & $\times$ &23,370\\
Google+~\cite{leskovec2014snap} &rumor& \checkmark & \checkmark & $\times$ &23,628\\
Email~\cite{leskovec2012learning} &rumor& \checkmark  & \checkmark & $\times$&1,005\\
Facebook~\cite{leskovec2007graph} &rumor& \checkmark & $\times$ & $\times$ &135\\
Twitter~\cite{lampos2014predicting} & voting & $\times$ & $\times$ & \checkmark &38,020\\
Twitter~\cite{rivadeneira2021predicting} &election & $\times$ & $\times$ & \checkmark &2\\
Twitter~\cite{lahuerta2016looking} &marketing& $\times$ & \checkmark & \checkmark &3,853\\
Twitter~\cite{mallipeddi2022framework} & marketing &  \checkmark & \checkmark & \checkmark & 37\\
\midrule
\textbf{SAGraph}&marketing& \checkmark & \checkmark & \checkmark &345,039\\
\bottomrule
\end{tabular}}
\label{table:data_comparision}
\end{table}

\begin{figure}[htb]
    \centering
    \includegraphics[width=0.8\linewidth]{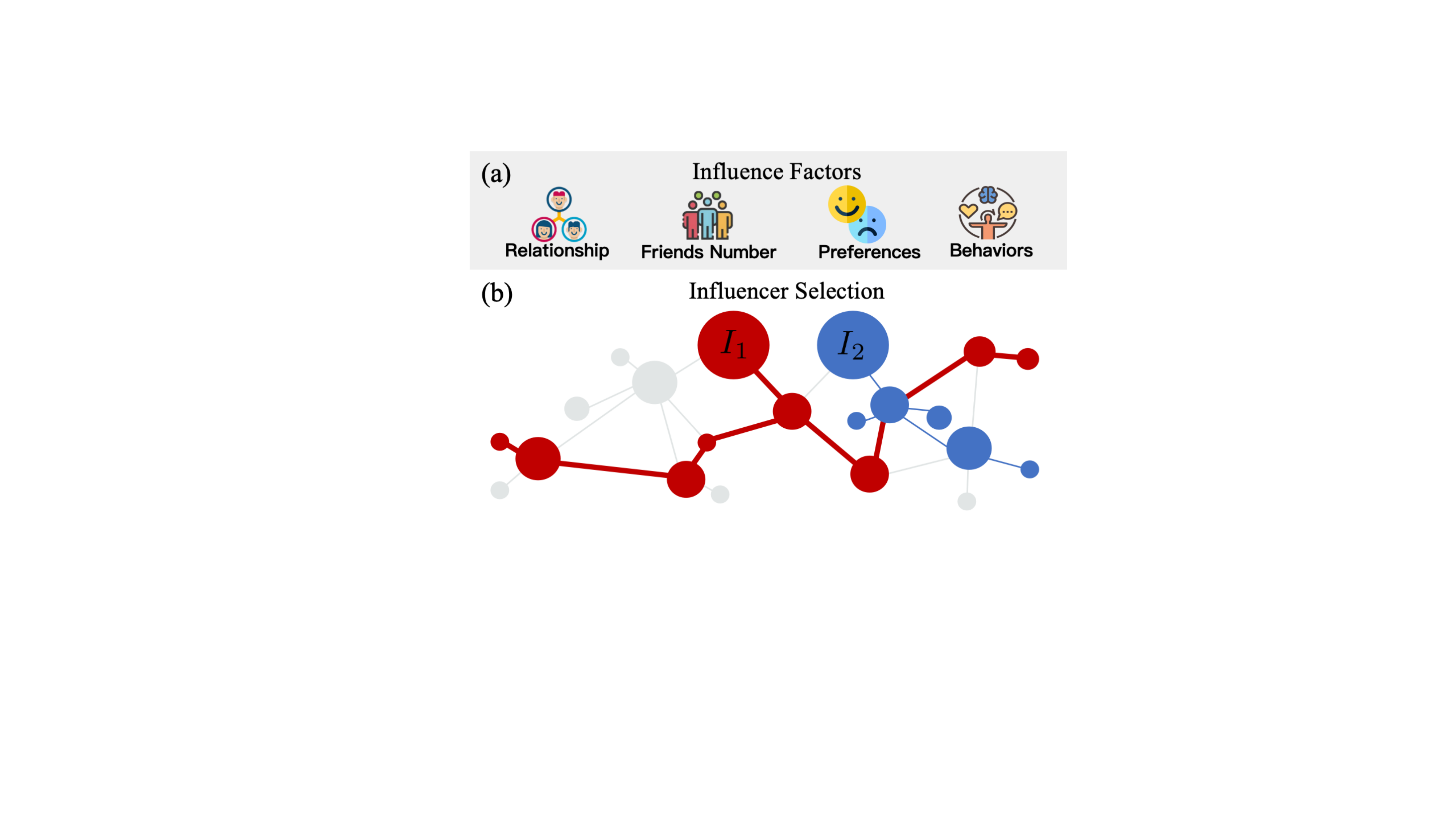}
    \caption{(a) An illustration of influence factors. (b) The process of influencer selection involves the information diffusion across various influence factors. ``$I_1$'' and ``$I_2$'' represent influencers selected due to their larger influence.
    }
    \label{fig:intro}
\end{figure}

\section{Introduction}
\label{intro}
Social networks and media platforms have revolutionized our interactions and information sharing, with influencers significantly impacting their followers' decisions through recommendations on products, services, and lifestyle choices.
In digital advertising, promoters strategically select a limited number of influencers to maximize their product's influence and visibility in the market~\cite{zhang2015influenced,mallipeddi2022framework,lenger2022choose,qiu2018deepinf}.
This approach benefits both advertisers and consumers alike.
As shown in Figure~\ref{fig:intro}(a), key factors in influencer selection include relationships, number of friends, preferences, and behaviors such as comments and reposts. Figure~\ref{fig:intro}(b) illustrates the information diffusion process, which leads influencers $I_1$ and $I_2$ to gain large influence due to their extensive reach.
Numerous studies have focused on influencer selection, leading to the development of datasets designed to enhance the effectiveness of influencer marketing strategies. For example, ~\citeauthor{lahuerta2016looking} analyzed 30,000 tweets from 3,853 users to explore how factors such as writing style, user sentiment, and follower count impact the calculation of influence. Additionally, ~\citeauthor{mallipeddi2022framework} constructed a Twitter dataset featuring profiles of 37 influencers who posted 18,571 tweets. By collecting retweet data, they built a social network to further investigate the effects of repeated exposure and the forgetting effect on influence calculation.

Expanding on this, datasets across various scenarios also concentrate on selecting important nodes within a graph, which is essentially similar to the influencer selection task.
For example, ~\citeauthor{leskovec2007cost} used extensive links to construct a social network of blogs. 
By analyzing the social network's connections, they found influential blogs for spreading information effectively.
Another widely used real-world dataset, SNAP\cite{leskovec2016snap}, includes Wiki Vote, Twitter Lists, and Google+ datasets. 
These datasets consist of directed social networks where nodes represent users and edges represent user interactions, such as voting on Wikipedia edits, following on Twitter, or adding users to circles on Google+.
All datasets are for influencer selection, particularly to combat rumors in social networks.

However, a significant issue with existing datasets is that they oversimplify complex social networks.
In marketing scenarios, for example, the network is often reduced by filtering tweets or user nodes based on keywords, which undermines the integrity of the social network. 
This oversimplification not only erases the intricate relationships within the network but also leads to misjudgments about the potential influence of individuals or content.
In non-marketing contexts, social networks are often distilled into numerical representations, ignoring the subtle semantic meanings in text and the emotional nuances of interactions.
People's opinions are nuanced and diverse, and the mechanisms of influence are intricate.
For example, the persuasive power of an influencer's message can be deeply intertwined with the language used, the context in which it is presented, and the emotional resonance it creates with the audience, aspects that are not fully captured by numerical metrics alone.

To address this gap, we present a comprehensive social advertising graph dataset, SAGraph, which includes multi-domain contextual data sourced from Weibo~\cite{yang2012automatic, li2020vmsmo, gao2019abstractive}, one of the most popular social networking platforms.
The dataset includes user information with domain-specific preferences such as interests, post data, and interaction context, all tailored for advertising and marketing scenarios.
Specifically, we begin with a set of ground truth influencers—product promoters—and a group of influencers who share the same domain and interests but are not product promoters.
For each user, we collect personal information such as friend numbers, bio details, posts, and interaction data, which includes comments and reposts.
Subsequently, we proceed iteratively to expand the graph.
In summary, our graph centers around 6 product campaigns in diverse domains, encompasses 345,039 users, each with their profile information, and published posts, and includes interaction data comprising 1,302,109 comments and 554,629 reposts.
In Table~\ref{table:data_comparision}, we compare our dataset with existing ones.
SAGraph is the first comprehensive social advertising graph dataset with detailed elements that integrate users, products, social relationships, interactions, and textual information to simulate real-world social environments. 
It significantly surpasses previous datasets in terms of scale, completeness, and domain diversity.

For evaluation, we experiment with several influencer selection baselines, including those in the simulation, proxy, sketch, and context-aware categories, as well as the latest LLMs on SAGraph.
The experiments show that classic baselines, which do not consider textual information, perform poorly on this task. 
Meanwhile, context-aware methods and LLMs show great potential in modeling user profiles and influencer prediction.

\begin{figure*}[htb]
  \centering
  \includegraphics[width=0.75\textwidth]{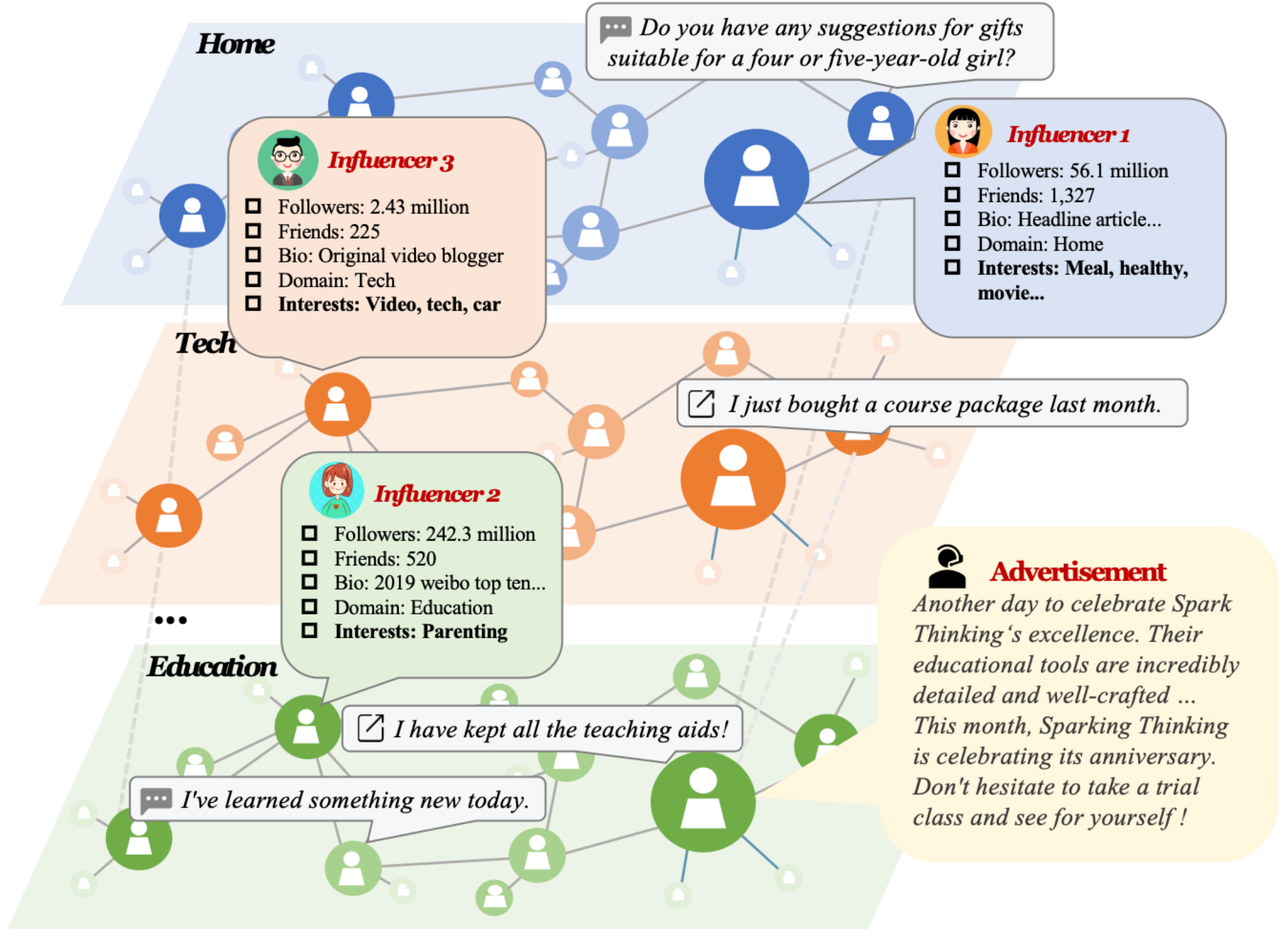}
  \caption{An example of the SAGraph, which includes users with interest tags, ads as posts, and evolving interaction data.}
  \label{fig:data}
\end{figure*}
The contribution of our work can be summarized as follows: 

$\bullet$ \textbf{A comprehensive social advertising graph dataset}: We introduce the large-scale SAGraph dataset, featuring domain-specific interest tags from Weibo, which is publicly available for advertising campaigns. This dataset includes a comprehensive graph structure and rich textual data from user profiles and interaction content, supporting the development of advanced algorithms for influencer selection and consumer behavior analysis.

$\bullet$ \textbf{Demonstration of benchmarks}: We evaluate influencer selection baselines on our dataset, showing text's role in influence calculation. For advertising, we offer an LLM-based tool to improve influencer selection using rich text, aiding strategic campaign decisions to increase brand visibility and sales. This underscores the dataset's practical value and potential for marketing research.

\section{Related Work}

\textbf{Influencer Selection Methods.}
Influencer selection methods have significant potential for applications in areas such as public opinion monitoring, advertising promotion, and epidemic prediction. 
Traditional simulation-based methods model information diffusion using a large number of Monte Carlo(MC) simulations to estimate the spread of influence. 
Notable methods include CELF~\cite{leskovec2007cost}, CELF++~\cite{goyal2011celf++}, and GREEDY~\cite{kempe2003maximizing}. While these methods are adaptable to various diffusion models, they come with a substantial computational cost.
To identify near-optimal seed sets more efficiently, proxy-based methods approximate the spread of influence using proxies like PageRank, shortest paths, and others. 
These methods generally have lower computational complexity and can deliver results in less time. 
Methods include EIGEN-CENTRALITY~\cite{zhong2018identifying}, DEGREE~\cite{Centrality}, SIGMA~\cite{yan2019minimizing}, and PI~\cite{zhang2022blocking}. 
However, these methods lack theoretical approximation guarantees and are highly sensitive to minor changes in network structure, often leading to unstable results.
Sketch-based methods, on the other hand, precompute ``sketches'' based on specific diffusion models and evaluate the spread of influence on these sketches, avoiding the need for repeated time-consuming MC simulations. 
While maintaining theoretical guarantees, these methods significantly improve computational efficiency. Representative examples include RIS~\cite{borgs2014maximizing}, TIM~\cite{tang2014influence}, IMM~\cite{tang2015influence}, and EaSyIM~\cite{galhotra2016holistic}.
Additionally, to better simulate real-world information diffusion and enhance result stability, context-aware methods integrate various contextual information such as topics, time, and location. 
This approach improves influencer selection performance by fully leveraging the potential of contextual data, enabling the development of more effective influencer selection algorithms~\cite{li2015real}.

\textbf{Influencer Selection Datasets.} Influencer selection datasets span multiple domains. In marketing, the Twitter dataset in ~\cite{lahuerta2016looking} is used to analyze the importance of text features for influence calculation but does not address the construction of a social network. Another Twitter dataset, which includes 37 influencers over a maximum period of 10 months, constructs a social network to examine the impact of multiple exposures and forgetting effects on influence calculation~\cite{mallipeddi2022framework}. However, the network is small, and tweets were filtered based on keywords, leading to missing information about the influencers and an incomplete social network for the specified group.
In non-marketing domains, datasets such as Blog~\cite{leskovec2007cost}, Tweet List~\cite{leskovec2014snap}, and Facebook~\cite{leskovec2007graph} are based on user-following relationships. By analyzing follower counts and the rate of content spread over a specific period, influencers can be predicted. For datasets containing interaction data, such as Wiki Vote~\cite{leskovec2014snap}, Email~\cite{leskovec2012learning}, Google+~\cite{leskovec2014snap}, and Twitter~\cite{mallipeddi2022framework}, the interactions are typically direct, involving information dissemination through targeted emails, knowledge sharing, and voting.
The study in ~\cite{kwak2010twitter} identifies influential users on Twitter by analyzing the timing and popularity of topic propagation. However, these datasets lack a comprehensive analysis of user content and fail to capture the nuances of text-based interactions and direct content feedback. As a result, they are inadequate for supporting in-depth research on the sources of influence, propagation factors, and propagation trajectories.


\section{Dataset Overview}
To our knowledge, there is no comprehensive benchmark dataset that includes advertisement, influencer, influencee, and interaction information within a social network. 
To fill this void, we compile such a dataset from Weibo, one of China's most widespread social media platforms. 
In this section, we initially outline the schema of the SAGraph, which encompasses advertising products, influencer details, and interaction data. Subsequently, we elucidate our collection strategy and finally, we describe the statistics of the data.

\begin{algorithm}[tb]
\small
\caption{SAGraph Collection}
\label{alg:idcalgorithm}
\begin{algorithmic}[1]
\State Initialize the seed user set $S$ with $k$ influencers' IDs who have more than 100k followers, $F$ as the set of influencees.
\State Set the user queue ${Q_U=S}$, the post queue $Q_P=\emptyset$, the whole collection rounds $N=6$, the current iteration $i=0$.
\While{ ${i<N}$ }
    \While{ $Q_U$ is not empty}
        \State Dequeue a user $u$ from $Q_U$.
        \State Collect the profile information of $u$.
        \State Fetch all the posts's IDs from $u's$ first page as $P_u$.
        \State Enqueue $P_u$ to the queue of posts $Q_P$.
    \EndWhile
    \While{ $Q_P$ is not empty}
        \State Dequeue a post $p$ from $Q_P$.
        \State Collect all interaction data under the post $p$ and select users with more than 100k followers as the new seed users  $U_p$.
        \State Enqueue $U_p$ without duplication back into the queue $Q_U$ for further data collection.
    \EndWhile
\EndWhile
\State We organize users, profile information, posts, and interaction data into our SAGraph dataset.
\label{idc}
\end{algorithmic}
\end{algorithm}

\subsection{Dataset Schema}
\noindent \textbf{Advertising products}. In our study, we identify and choose a trending product within each of the six domains—education, technology, hygiene, skincare, linguistics, and home—for promotion.
These products include `Spark Thinking', an educational platform that fosters creativity and learning; the `Intelligent Floor Scrubber', representing advanced cleaning technology; the `Electric Toothbrush', used for oral hygiene;  the `Ruby Face Cream', for daily skincare; the `ABCReading', for encouraging reading and communication for language enhancement, and the `SUPOR Boosted Showerhead' for providing an invigorating shower experience at home. 
Each product is chosen for its distinct appeal and market potential. 
The advertising data encompasses the names of the promoted products, their respective domains, and the promotional copy.
We show the translated version of all content in this paper.

\noindent \textbf{Influencer details}.
For each influencer, we collect their profiles, which include friend numbers, bio information, and the most recent posts as part of the influencer details. 
The number of friends is included to initially estimate the exposure range of influencer marketing campaigns.
Biographical details illuminate the influencer's areas of interest and expertise, aligning them with their audience's interests.
Recent posts reflect current engagement topics, enabling timely and targeted influencer strategies.
For influencers in each domain and their interacting influencees, we utilize the profiles and historical interaction content to invoke GPT-4(gpt-4-1106-preview), generating preferences such as interest tags for each user.
This text-rich dataset refines our selection of influencers for effective marketing initiatives.
Our influencers include at least four individuals who are promoters as well as candidate influencers.

\noindent \textbf{Interaction data}. 
We collect the interaction data consisting of reposts and comments originating from posts. Just like comments, reposts also include text content that expresses opinions. Figure \ref{fig:data} illustrates the interaction detail of each influencer. 
For each microblog post, we collect comprehensive interaction information, which includes not only direct interactions such as user comments and reposts but also multiple interactions between users under the host's post and consecutive multi-round interactions of users. This extensive data engagement guarantees a more holistic and multifaceted perspective on user interactions. Finally, we get a detailed list of users interacting with the influencers.

\begin{table*}[t]
  \centering
  \caption{Statistics of SAGraph across construction. `R' represents the collection round.}
   \resizebox{\linewidth}{!}{
\begin{tabular}{cccccccccccccc}
\toprule
\multirow{2}[4]{*}{\textbf{Product}} & \multicolumn{6}{c}{\textbf{\#Infuencers}}     & \multirow{2}[4]{*}{\textbf{\#User}} & \multirow{2}[4]{*}{\textbf{\#Interaction}} & \multirow{2}[4]{*}{\textbf{\#Comment}} & \multirow{2}[4]{*}{\textbf{\#Repost}} & \multicolumn{1}{c}{\multirow{2}[4]{*}{\textbf{\makecell{\#Interaction \\ Per \#Follower}}}} & \multicolumn{1}{c}{\multirow{2}[4]{*}{\textbf{\makecell{Variance \\ of \#Followers}}}} & \multicolumn{1}{c}{\multirow{2}[4]{*}{\textbf{\makecell{Variance \\ of \#Interactions}}}} \\
\cmidrule{2-7}      & \textbf{R1} & \textbf{R2} & \textbf{R3} & \textbf{R4} & \textbf{R5} & \textbf{R6} &       &       &       &       &       &       &  \\
\midrule
\textbf{Electric Toothbrush} & 10    & 17    & 26    & 44    & 46    & 48    & 43,611 & 163,016 & 88,214 & 74,804 & 3.73  & 120.12 & 97.9 \\
\textbf{Spark Thinking} & 10    & 11    & 17    & 35    & 88    & 201   & 58,510 & 490,431 & 390,718 & 99,713 & 8.38  & 167.17 & 134.66 \\
\textbf{Intelligent Floor Scrubber} & 10    & 20    & 27    & 58    & 98    & 190   & 63,619 & 370,176 & 269,769 & 100,407 & 5.82  & 200.21 & 130.35 \\
\textbf{Ruby Face Cream} & 10    & 11    & 12    & 14    & 19    & 55    & 66,248 & 126,859 & 76,349 & 50,510 & 1.91  & 132.65 & 108.3 \\
\textbf{ABCReading} & 10    & 14    & 25    & 41    & 90    & 206   & 70,258 & 458,513 & 343,882 & 114,631 & 6.52  & 225.16 & 149.1 \\
\textbf{SUPOR Boosted Showerhead} & 10    & 27    & 31    & 58    & 74    & 81    & 138,427 & 247,741 & 133,177 & 114,564 & 1.79  & 102.25 & 98.61 \\
\bottomrule
\end{tabular}%
}%
   \label{table:statistic}%
\end{table*}%

\begin{table*}[t]
  \centering
  \caption{Statistics of SAGraph across six domains.}
   \resizebox{\linewidth}{!}{
   \begin{tabular}{cccccccccc}
   \hline
   \textbf{Product} &\textbf{Domain}& \textbf{\#User} & \textbf{\makecell{\#Tagged \\User}} & \textbf{\#Tags} & \textbf{\makecell{\#Duplicated \\Tags}} & \textbf{\#Influencers} & \textbf{\makecell{\#Posts of \\Influencers}} & \textbf{\makecell{\#Prompted \\Influencers}} & \textbf{\#Label Examples} \\
   \hline
\textbf{Electric Toothbrush} & Hygiene & 43,611 & 22,675 & 50,990 & 9,427 & 48    & 1,766 & 8     & Healthy,Cleaning \\
\textbf{Spark Thinking} & Education & 58,510 & 27,413 & 74,831 & 11,311 & 201   & 34,058 & 5     & Education,Parenting,Mathematics \\
\textbf{Intelligent Floor Scrubber} & Technology & 63,619 & 33,025 & 85,734 & 13,025 & 190   & 2,490 & 5     & Digital,Technology \\
\textbf{Ruby Face Cream} & Skincare & 66,248 & 27,257 & 58,289 & 9,488 & 55    & 844   & 14    & Beauty,Skincare,Makeup,Shopping \\
\textbf{ABCReading} & Linguistics & 70,258 & 34,341 & 88,546 & 13,107 & 206   & 3,176 & 4     & Language,English \\
\textbf{SUPOR Boosted Showerhead} & Home  & 138,427 & 121,649 & 228,761 & 19,050 & 81    & 1,848 & 6     & Home,Daily Life,Furniture \\
\hline
   \end{tabular}%
   }
   \label{table:labels}%
\end{table*}

\subsection{Dataset Construction}
\label{collection}
\subsubsection{Data Collection}
Social networks based on follow-up relationships tend to be sparse, with the presence of inactive users and fake accounts.
To enrich our network with more meaningful interaction data, we incorporate comments and reposts into the social network construction process.
First, we select at least four bloggers who have promoted the assigned products. 
Utilizing the Weibo platform, which regularly offers a curated list of related users for each account—based on shared interests, follower overlap, and similar domains—we proceed to choose additional users from this list who have not engaged in product promotion. 
These selected users, both promoters and non-promoters, serve as our initial seed users, totaling ten.
To ensure fairness, we select non-promoters with a follower count comparable to that of the promoters.
Next, we collect all posts from these seed users’ pages and identify users who have commented or reposted from the interaction data under the posts to construct the nodes and edges for the social network. 
To ensure continuous expansion of the social network, we iterate this process six times, adhering to the Six Degrees of Separation principle~\cite{guare2016six}.
Each iteration we seek new seed users to expand our collection of posts and interaction data. 
We specifically target users with more than 100k followers within the already-established social network to serve as our new seed users.
In our data collection experiment, we set the threshold at 100k followers to balance the recall of users within the target domain while preventing the exponential growth of the social network, which could disrupt the data collection process. By setting the 100k threshold, we ensure that the data expansion remains stable, with each iteration growing by no more than an order of magnitude of 10.
The overall process is shown in Social Advertisement Graph Collection Algorithm~\ref{alg:idcalgorithm}.
SAGraph is released under the Creative Commons
CC-BY 4.0 license that academics and industry may use.

\subsubsection{Dataset Anonymization}
We anonymize the collected user information to ensure the protection of user privacy. 
First, we transformed potentially revealing usernames and real names by assigning each user a new ID and username. 
Next, we filtered sensitive fields such as regions, locations, and gender. 
In interaction content, we replaced the original usernames with the transformed ones and used special placeholders to handle sensitive information, such as phone numbers. 
These measures effectively conceal users' real information while preserving the original interaction relationships through the generated user IDs and usernames. 
For detailed processing steps, please refer to our GitHub repository.

\subsubsection{Dataset Ethics}
Our data construction adheres to Chapter III, Process and Use of Data, from the Measures for Data Security Management~\cite{datasecuritymanagement2019}. We only scrape content that is publicly accessible and ensure its anonymization by removing all personally identifiable information.
Compared to previous work on Weibo data ~\cite{zhang2013social,li2018retweeting,zhang2015influenced,cao2020popularity,li2020data}, our processing procedures are more stringent.
We have also provided the complete code, including data collection and anonymization process, allowing anyone to replicate the acquisition of the SAGraph dataset easily.
Additionally, we will provide the original data upon request, subject to a thorough review and upon signing a consent agreement, ensuring compliance with our data use policies.

\begin{figure}[htb]
    \centering
    \includegraphics[width=0.8\linewidth]{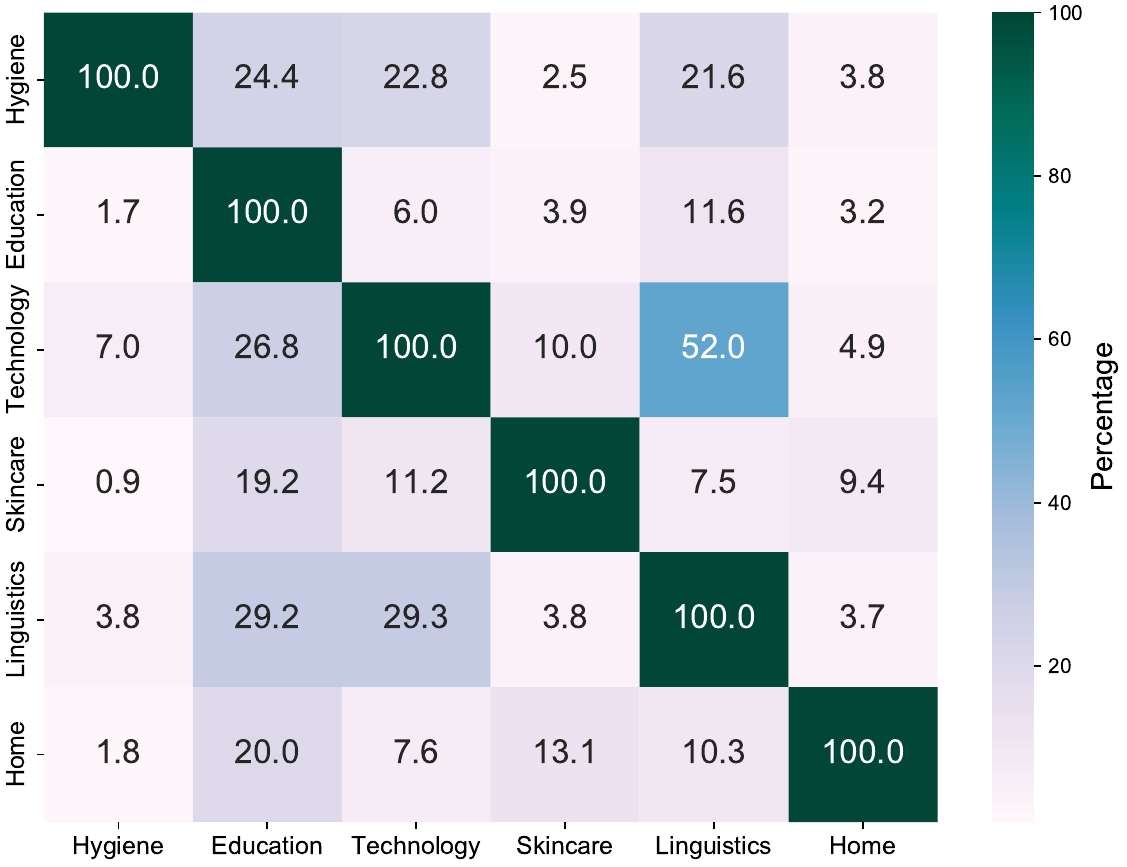}
    \caption{The percentage of user overlap across domains.
    }
    \label{fig:overlap}
\end{figure}

\subsection{Dataset Statistics}
The graph dataset statistics for each advertising product are shown in Table~\ref{table:statistic}.
During the iterative process, noticeable variations are observed in the rate of user expansion across different domains. 
Products like the Electric Toothbrush, Ruby Face Cream, and SUPOR Boosted Showerhead exhibit a steady rise in new seed users. 
While domains like ABCReading, Spark Thinking, and Intelligent Floor Scrubber experience a pronounced escalation in the recruitment of seed users, showing a trend of doubling with each successive iteration.
This trend aligns with the interaction metrics of these domains. The former group has a lower variance of 101.6 in user interactions, indicating stable and focused engagement. 
In contrast, the latter group displays higher interaction rounds of 6 per follower and a larger variance of 138, suggesting a dynamic and swiftly growing user base. 
These observations highlight the intricate dynamics of user engagement and expansion across various markets.

The statistics for the domain data are presented in Table \ref{table:labels} and Figure \ref{fig:overlap}. Table \ref{table:labels} offers a detailed overview of each domain, including the number of users, the number of users with interest tags, the total number of tags, the number of duplicated tags, the number of influencers, the number of posts made by influencers, and the number of influencers who have conducted advertising promotions within each domain. 
We also list the typical user labels for each domain. 
Approximately 60\% of users successfully generated interest tags. 
Users without interest tags either lacked personalized signatures or did not engage in interactions such as comments or reposts, thus failing to provide sufficient data for the generation of interest tags. 
On average, each influencer published over a hundred posts, with a series of interactions forming around these posts, providing abundant information for user behavior analysis.
Figure \ref{fig:overlap} illustrates the overlap of users across the six domains. 
The lower triangle values represent the proportion of overlapping users in the domain on the horizontal axis, while the upper triangle values represent the proportion of overlapping users in the domain on the vertical axis. 
Clear differences are observed among users in different domains, with an average overlap rate of 18\%. 
However, some commonalities also emerge. 
For instance, overlapping users from all domains represent a notable proportion in the education domain (close to 20\%), while the hygiene domain shows the least overlap with other domains (around 3\%). The highest overlap is seen between the linguistics and technology domains, at 52\%.

\begin{table*}[t!]
  \centering
     \caption{Performance comparison of simulation baselines and proxy baselines on six datasets.}
  \resizebox{0.9\linewidth}{!}{
\begin{tabular}{clrrrrrrrrrrrr}
\toprule
\multicolumn{2}{c}{\multirow{2}[4]{*}{\textbf{Models}}} & \multicolumn{6}{c}{\textbf{Electric Toothbrush}} & \multicolumn{6}{c}{\textbf{Spark Thinking}} \\
\cmidrule{3-14}\multicolumn{2}{c}{} & \multicolumn{1}{l}{\textbf{ P@5 }} & \multicolumn{1}{l}{\textbf{ P@10 }} & \multicolumn{1}{l}{\textbf{ R@5 }} & \multicolumn{1}{l}{\textbf{ R@10 }} & \multicolumn{1}{l}{\textbf{ G@5 }} & \multicolumn{1}{l}{\textbf{ G@10 }} & \multicolumn{1}{l}{\textbf{ P@5 }} & \multicolumn{1}{l}{\textbf{ P@10 }} & \multicolumn{1}{l}{\textbf{ R@5 }} & \multicolumn{1}{l}{\textbf{ R@10 }} & \multicolumn{1}{l}{\textbf{ G@5 }} & \multicolumn{1}{l}{\textbf{ G@10}} \\
\midrule
\multirow{3}[2]{*}{\textbf{Simulation-based}} & \textbf{CELF  } & 0.40   & 0.40   & 0.40   & 0.80   & 0.51  & 0.73  & 0.20   & 0.20   & 0.25  & 0.50   & 0.15  & 0.28 \\
      & \textbf{CELF++  } & 0.40   & 0.20   & 0.40   & 0.40   & 0.32  & 0.32  & 0.40   & 0.20   & 0.50   & 0.50   & 0.64  & 0.64 \\
      & \textbf{GREEDY} & 0.20   & 0.30   & 0.20   & 0.60   & 0.15  & 0.37  & 0.40   & 0.20   & 0.50   & 0.50   & 0.64  & 0.64 \\
\midrule
\multirow{4}[2]{*}{\textbf{Proxy-based}} & \textbf{EIGEN-CENTRALITY} & 0.40   & 0.40   & 0.40   & 0.80   & 0.36  & 0.60   & 0.20   & 0.20   & 0.25  & 0.50   & 0.17  & 0.29 \\
      & \textbf{DEGREE} & 0.40   & 0.40   & 0.40   & 0.80   & 0.51  & 0.74  & 0.20   & 0.30   & 0.25  & 0.75  & 0.25  & 0.48 \\
      & \textbf{SIGMA  } & 0.20   & 0.30   & 0.20   & 0.60   & 0.15  & 0.37  & 0.00     & 0.20   & 0.00     & 0.5   & 0.00     & 0.23 \\
      & \textbf{PI  } & 0.40   & 0.30   & 0.40   & 0.60   & 0.37  & 0.46  & 0.00     & 0.20   & 0.00     & 0.50   & 0.00     & 0.23 \\
\midrule
      &       & \multicolumn{6}{c}{\textbf{Ruby Face Cream}}  & \multicolumn{6}{c}{\textbf{Intelligent Floor Scrubber}} \\
\midrule
\multirow{3}[2]{*}{\textbf{Simulation-based}} & \textbf{CELF  } & 0.40   & 0.40   & 0.29  & 0.57  & 0.35  & 0.46  & 0.00     & 0.20   & 0.00     & 0.50   & 0.00     & 0.26 \\
      & \textbf{CELF++  } & 0.40   & 0.20   & 0.29  & 0.29  & 0.51  & 0.41  & 0.40   & 0.20   & 0.50   & 0.50   & 0.64  & 0.64 \\
      & \textbf{GREEDY} & 0.40   & 0.20   & 0.29  & 0.29  & 0.51  & 0.41  & 0.40   & 0.20   & 0.50   & 0.50   & 0.44  & 0.44 \\
\midrule
\multirow{4}[2]{*}{\textbf{Proxy-based}} & \textbf{EIGEN-CENTRALITY} & 0.60   & 0.50   & 0.43  & 0.71  & 0.62  & 0.68  & 0.20   & 0.20   & 0.25  & 0.50   & 0.17  & 0.17 \\
      & \textbf{DEGREE} & 0.80   & 0.50   & 0.57  & 0.71  & 0.79  & 0.72  & 0.20   & 0.30   & 0.25  & 0.75  & 0.15  & 0.40 \\
      & \textbf{SIGMA  } & 0.60   & 0.40   & 0.43  & 0.57  & 0.53  & 0.53  & 0.20   & 0.20   & 0.25  & 0.50   & 0.25  & 0.36 \\
      & \textbf{PI  } & 0.60   & 0.40   & 0.43  & 0.57  & 0.66  & 0.63  & 0.20   & 0.20   & 0.25  & 0.50   & 0.25  & 0.36 \\
\midrule
      &       & \multicolumn{6}{c}{\textbf{ABC Reading}}      & \multicolumn{6}{c}{\textbf{SUPOR Boosted Showerhead}} \\
\midrule
\multirow{3}[2]{*}{\textbf{Simulation-based}} & \textbf{CELF  } & 0.20   & 0.20   & 0.25  & 0.50   & 0.39  & 0.50   & 0.20   & 0.20   & 0.25  & 0.50   & 0.20   & 0.33 \\
      & \textbf{CELF++ } & 0.40   & 0.30   & 0.50   & 0.75  & 0.35  & 0.49  & 0.20   & 0.10   & 0.25  & 0.25  & 0.25  & 0.25 \\
      & \textbf{GREEDY} & 0.20   & 0.10   & 0.25  & 0.25  & 0.39  & 0.39  & 0.40   & 0.20   & 0.50   & 0.50   & 0.64  & 0.64 \\
\midrule
\multirow{4}[2]{*}{\textbf{Proxy-based}} & \textbf{EIGEN-CENTRALITY} & 0.20   & 0.30   & 0.25  & 0.75  & 0.15  & 0.41  & 0.00     & 0.10   & 0.00     & 0.25  & 0.00     & 0.11 \\
      & \textbf{DEGREE} & 0.00     & 0.20   & 0.00     & 0.50   & 0.00     & 0.25  & 0.00     & 0.10   & 0.00     & 0.25  & 0.00     & 0.11 \\
      & \textbf{SIGMA  } & 0.00     & 0.20   & 0.00     & 0.50   & 0.00     & 0.25  & 0.20   & 0.10   & 0.25  & 0.25  & 0.39  & 0.39 \\
      & \textbf{PI  } & 0.00     & 0.20   & 0.00     & 0.50   & 0.00     & 0.25  & 0.20   & 0.10   & 0.25  & 0.25  & 0.39  & 0.39 \\
\bottomrule
\end{tabular}%
}%
     \label{tab:main}%
\end{table*}%

\section{Task Definition}
Before introducing our dataset construction process, we first give a formal definition of the influencer selection task.
We denote a social network as $G=(V, E)$.
Here, each node $v_i$ corresponds to a user, including personal details such as friends number and bio information. 
$V$ is divided into two distinct categories: one is defined as $V_1$ that comprises normal users with fewer than 100K followers, and the other is defined as $V_2$ that consists of seed users with 100K followers or more.
The edges $E$ denote various interactions between users, including commenting and reposting content, along with the associated original post content. 
Multiple edges can exist between users if there are repeated interactions, such as a follower commenting on different posts by the same blogger.
For a product that requires promotion, our objective is to identify a list of influencers from $V_2$, specifically those with the greatest potential to \textit{influence} their influencees. 
In our framework, \textit{influence} is defined as the ability of an influencer to increase the likelihood of a follower purchasing an advertised product by eliciting a stronger intent to purchase through their actions or content.

\section{Experimental Studies}

\begin{table*}[t]
  \centering
  \caption{Performance of sketch-based and context-based methods on six datasets.}
  \small

\begin{tabular}{lrrrrrrrrrrrr}
\toprule
\multicolumn{1}{c}{\multirow{2}[4]{*}{\textbf{Models}}} & \multicolumn{6}{c}{\textbf{Electric Toothbrush}} & \multicolumn{6}{c}{\textbf{Spark Thinking}} \\
\cmidrule{2-13}      & \multicolumn{1}{l}{\textbf{ P@5 }} & \multicolumn{1}{l}{\textbf{ P@10 }} & \multicolumn{1}{l}{\textbf{ R@5 }} & \multicolumn{1}{l}{\textbf{ R@10 }} & \multicolumn{1}{l}{\textbf{ G@5 }} & \multicolumn{1}{l}{\textbf{ G@10 }} & \multicolumn{1}{l}{\textbf{ P@5 }} & \multicolumn{1}{l}{\textbf{ P@10 }} & \multicolumn{1}{l}{\textbf{ R@5 }} & \multicolumn{1}{l}{\textbf{ R@10 }} & \multicolumn{1}{l}{\textbf{ G@5 }} & \multicolumn{1}{l}{\textbf{ G@10}} \\
\midrule
\textbf{RIS} & 0.40   & 0.30   & 0.40   & 0.60   & 0.36  & 0.47  & 0.00     & 0.10   & 0.00     & 0.25  & 0.00     & 0.12 \\
\textbf{KB-TIM} & 0.20   & 0.30   & 0.20   & 0.60   & 0.21  & 0.56  & 0.40   & 0.30   & 0.50   & 0.75  & 0.35  & 0.60 \\
\midrule
      & \multicolumn{6}{c}{\textbf{Ruby Face Cream}}  & \multicolumn{6}{c}{\textbf{Intelligent Floor Scrubber}} \\
\midrule
\textbf{RIS} & 0.60   & 0.40   & 0.43  & 0.57  & 0.53  & 0.52  & 0.20   & 0.30   & 0.25  & 0.75  & 0.20   & 0.43 \\
\textbf{KB-TIM} & 0.60   & 0.40   & 0.43  & 0.57  & 0.53  & 0.61  & 0.40   & 0.30   & 0.50   & 0.75  & 0.41  & 0.67 \\
\midrule
      & \multicolumn{6}{c}{\textbf{ABC Reading}}      & \multicolumn{6}{c}{\textbf{SUPOR Boosted Showerhead}} \\
\midrule
\textbf{RIS} & 0.20   & 0.20   & 0.25  & 0.50   & 0.15  & 0.26  & 0.20   & 0.10   & 0.25  & 0.25  & 0.39  & 0.39 \\
\textbf{KB-TIM} & 0.20   & 0.30   & 0.25  & 0.75  & 0.20   & 0.45  & 0.20   & 0.20   & 0.25  & 0.50   & 0.17  & 0.42 \\
\bottomrule
\end{tabular}%
   \label{table:context}%
\end{table*}%

\subsection{Baselines and Test Settings}
\subsubsection{Baselines}
We first evaluate our dataset with several classic influencer selection models. 
For \textit{simulation} baselines which estimate the influence of specific nodes by modeling the spread of information, we include \textbf{CELF}~\cite{leskovec2007cost}, \textbf{CELF++}~\cite{goyal2011celf++} and \textbf{GREEDY}~\cite{kempe2003maximizing}.
For \textit{proxy} baselines which use simplified models for approximating influence to reduce computational complexity, we include \textbf{EIGEN-CENTRALITY}~\cite{zhong2018identifying}, \textbf{DEGREE}~\cite{Centrality}, \textbf{SIGMA}~\cite{yan2019minimizing}, and \textbf{PI}~\cite{zhang2022blocking}.
For \textit{sketch} baselines we experiment \textbf{RIS}~\cite{borgs2014maximizing} for efficient influence estimation in networks.

We also employed the context-based method \textbf{KB-TIM}~\cite{li2015real} to evaluate our dataset, which incorporates the topic of users into the calculation of influence propagation.
Given the significant advantages of LLMs in text processing and social network simulation, we further used GPT-4 and Kimi~\footnote{https://platform.moonshot.cn} to model influence propagation and compared it with the following three types of baselines: \noindent $\bullet$ \textbf{GPT-4} serves as the LLM baseline. For each influencer candidate, we simulate their posting of advertisements and model the behavior of users within their social network. The behaviors include ignoring, commenting, and reposting. Each generated comment is evaluated with a purchase likelihood score (on a scale of 0-10) using GPT-4. Finally, we aggregate the purchase scores of all simulated comments for each advertisement to determine the overall purchase inclination, which serves as the measure of the influencer's influence. \noindent $\bullet$ \textbf{GPT-4 w/ profile} integrates the interests of influencers and users into the prompt for simulating comments, thereby enhancing the realism and relevance of the generated content. \noindent $\bullet$ \textbf{GPT-4 w/ profile\&CoT} incorporates an additional step-by-step reasoning process into the comment simulation. It mimics how influencees behave in the real world, with the goal of further improving the effectiveness of the simulated comment generation.

\subsubsection{Test Settings}
In the task of advertising campaigns, we have the IDs of real promoters. At the same time, we employed various baselines for comparison. With the budget set to 10, we compared the 10 most influential users identified by the baselines to the ground truth.
During the graph construction process, we used interaction heat as the activation weight, calculated by dividing the historical interaction frequency by the highest interaction frequency in the network. For information diffusion, all classic methods employed the Independent Cascade Model~\cite{kempe2003maximizing} for simulation, with the number of diffusion rounds set to 100 for each model.
We adopt standard Top-$k$ ranking metrics to assess the quality of our recommendations. These metrics include Precision ($P@k$), which measures the proportion of relevant items within the top-$k$ recommendations, Recall ($R@k$), which evaluates how many of the total relevant items are captured in the top-k recommendations, and Normalized Discounted Cumulative Gain ($G@k$), which assesses the ranking quality by valuing correct recommendations that appear higher in the list more favorably. 

\begin{table*}[t]
  \centering
  \caption{Performance of LLM-based methods on six datasets.}
  \small
\begin{tabular}{lrrrrrrrrrrrr}
\toprule
\multicolumn{1}{c}{\multirow{2}[4]{*}{\textbf{Models}}} & \multicolumn{6}{c}{\textbf{Electric Toothbrush}} & \multicolumn{6}{c}{\textbf{Spark Thinking}} \\
\cmidrule{2-13}      & \multicolumn{1}{l}{\textbf{ P@5 }} & \multicolumn{1}{l}{\textbf{ P@10 }} & \multicolumn{1}{l}{\textbf{ R@5 }} & \multicolumn{1}{l}{\textbf{ R@10 }} & \multicolumn{1}{l}{\textbf{ G@5 }} & \multicolumn{1}{l}{\textbf{ G@10 }} & \multicolumn{1}{l}{\textbf{ P@5 }} & \multicolumn{1}{l}{\textbf{ P@10 }} & \multicolumn{1}{l}{\textbf{ R@5 }} & \multicolumn{1}{l}{\textbf{ R@10 }} & \multicolumn{1}{l}{\textbf{ G@5 }} & \multicolumn{1}{l}{\textbf{ G@10}} \\
\midrule
\textbf{GPT-4 } & 0.40   & 0.40   & 0.40   & 0.80   & 0.36  & 0.60   & 0.40   & 0.20   & 0.50   & 0.50   & 0.35  & 0.35 \\
\textbf{GPT-4 w/ Profile } & 0.40   & 0.40   & 0.40   & 0.80   & 0.49  & 0.71  & 0.40   & 0.30   & 0.50   & 0.75  & 0.41  & 0.54 \\
\textbf{GPT-4 w/ Profile\&CoT} & 0.40   & 0.40   & 0.40   & 0.80   & 0.55  & 0.77  & 0.40   & 0.30   & 0.50   & 0.75  & 0.56  & 0.68 \\
\midrule
      & \multicolumn{6}{c}{\textbf{Ruby Face Cream}}  & \multicolumn{6}{c}{\textbf{Intelligent Floor Scrubber}} \\
\midrule
\textbf{GPT-4 } & 0.60   & 0.50   & 0.43  & 0.71  & 0.53  & 0.61  & 0.20   & 0.20   & 0.25  & 0.50   & 0.39  & 0.51 \\
\textbf{GPT-4 w/ Profile } & 0.40   & 0.40   & 0.29  & 0.57  & 0.38  & 0.49  & 0.40   & 0.30   & 0.50   & 0.75  & 0.54  & 0.68 \\
\textbf{GPT-4 w/ Profile\&CoT} & 0.60   & 0.50   & 0.43  & 0.71  & 0.53  & 0.60   & 0.40   & 0.30   & 0.50   & 0.75  & 0.59  & 0.72 \\
\midrule
      & \multicolumn{6}{c}{\textbf{ABC Reading}}      & \multicolumn{6}{c}{\textbf{SUPOR Boosted Showerhead}} \\
\midrule
\textbf{GPT-4 } & 0.40   & 0.20   & 0.50   & 0.50   & 0.32  & 0.32  & 0.40   & 0.20   & 0.50   & 0.50   & 0.40   & 0.40 \\
\textbf{GPT-4 w/ Profile  } & 0.40   & 0.20   & 0.50   & 0.50   & 0.36  & 0.36  & 0.40   & 0.20   & 0.50   & 0.50   & 0.49  & 0.40 \\
\textbf{GPT-4 w/ Profile\&CoT} & 0.40   & 0.30   & 0.50   & 0.75  & 0.64  & 0.75  & 0.40   & 0.20   & 0.50   & 0.50   & 0.40   & 0.40 \\
\bottomrule
\end{tabular}%
   \label{table:llm}%
\end{table*}%

\subsection{Baseline Results and Analysis}
We show the overall performance of all baselines in Table~\ref{tab:main},~\ref{table:context}, and ~\ref{table:llm}.

\subsubsection{Simulation-based}. The results of simulation-based methods in Table~\ref{tab:main} show relatively consistent performance across different products. While the recall and accuracy metrics are generally low, the NDCG metric is higher in some domains, suggesting that within the selected influencers, the ground truth rankings are relatively strong. However, these methods do not perform equally well for all products. GREEDY demonstrates better stability compared to CELF++, while CELF performs the worst overall. Interestingly, CELF performs the best for the ``Electric Toothbrush''. This discrepancy may be due to differences in user behavior and data distribution across products. For example, the user interaction patterns for certain products, such as the ``Electric Toothbrush'', may be more aligned with the CELF algorithm, while other products might be better suited to GREEDY or CELF++.

\subsubsection{Proxy-based}. Proxy-based methods exhibit similar performance characteristics to simulation-based methods, displaying notable instability. This instability can be quite pronounced. For instance, DEGREE performs well on several products but shows extremely low NDCG metrics for ``ABC Reading'' and ``SUPOR Boosted Showerhead''. Among the methods compared, SIGMA and PI demonstrate the best stability, while EIGEN-CENTRALITY exhibits the worst stability. In terms of average performance, DEGREE stands out as the best-performing method.

\subsubsection{Sketch-based}. Compared to simulation-based and proxy-based methods, sketch-based methods such as RIS in Table~\ref{table:context} demonstrate the most stable performance, with real promoters consistently appearing in the Top 5 results. However, its average performance is relatively weaker. This phenomenon may be linked to the implementation mechanism of RIS. RIS relies on random sampling to estimate node influence. While this method theoretically can cover key nodes in the network, in practice, the sampling process may introduce biases. Specifically, when using interaction frequency between users as sampling weights, the model may overly focus on users with high interaction rates, while neglecting the preference characteristics between users. This bias could prevent the model from fully capturing the diverse patterns of influence propagation within the network, thereby affecting its overall performance.

\subsubsection{Context-based}. In Table \ref{table:context}, the context-based method KB-TIM builds upon RIS by incorporating user interest information. The results show that KB-TIM significantly improves average performance across various recommendation metrics for the six products compared to RIS, while maintaining the stability of RIS. This phenomenon further emphasizes the importance of textual information in calculating influence.

\subsubsection{LLM-based}
\begin{figure}[htb]
    \centering
    \includegraphics[width=0.89\linewidth]{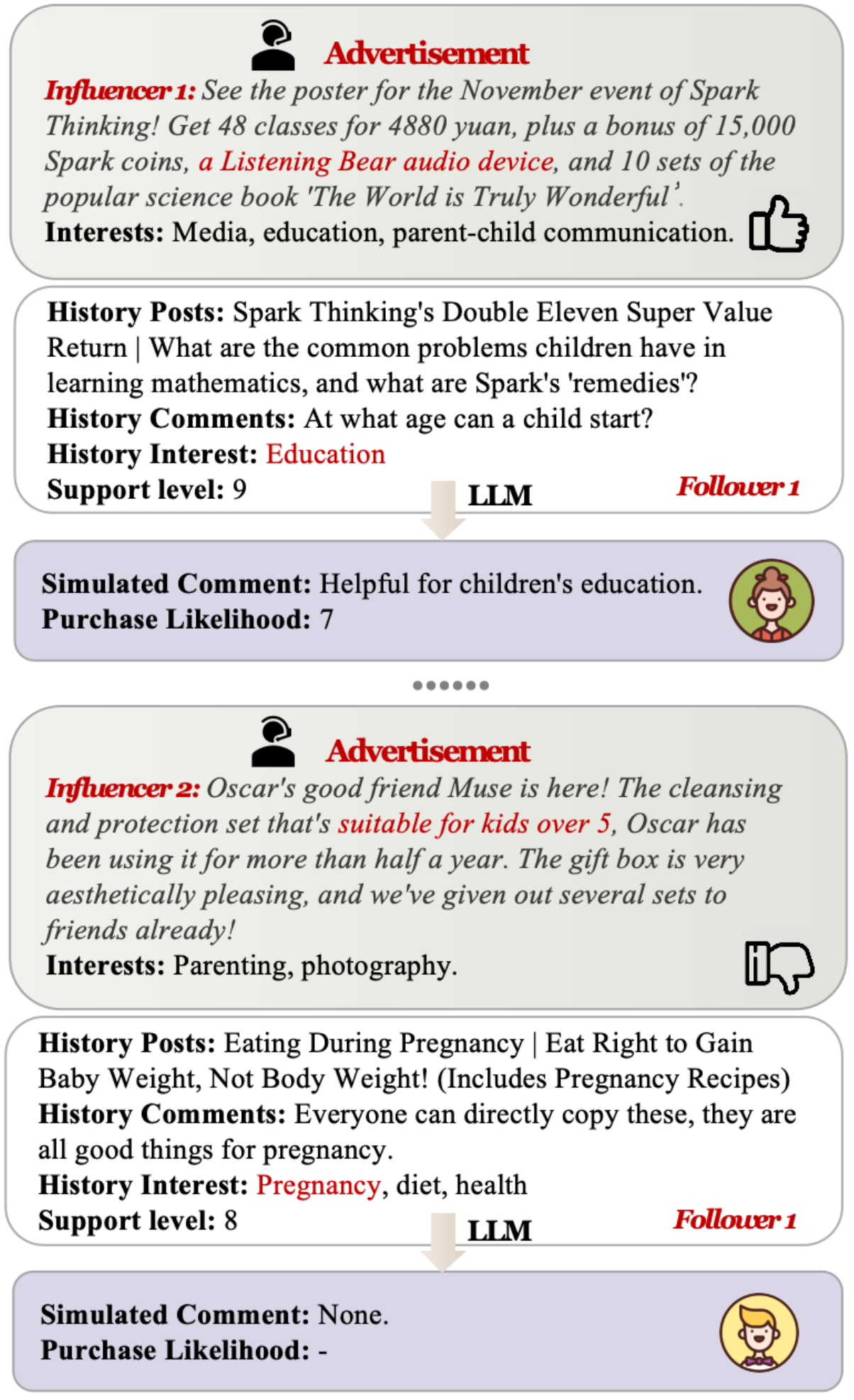}
    \caption{The comparison of influencers under LLM's simulation: The ``comment'' action is associated with a purchase likelihood, while the ``ignore'' action outputs ``None''.
    }
    \label{fig:show_case}
\end{figure}

GPT-4 outperforms all classic influencer selection baselines in terms of average performance at G@10, showing a significant improvement over the consistently stable performance of KB-TIM.
However, GPT-4 does not excel in terms of optimal performance across individual datasets. This is attributed to the LLMs' tendency to produce indistinguishable outputs in the absence of specific expressive personas.
Secondly, GPT-4 significantly outperforms all the classic influencer selection baselines when augmented with the profile and CoT modules.
By leveraging our user profile generation module, the LLM generates highly matched statements for the user population. With the CoT execution, the precision of the outputs increases, leading to a continuous enhancement in overall performance. This improvement is not only reflected in the significant boost in average performance but also in achieving the best performance on all the datasets except Ruby Face Cream.
The underperformance of the classic baselines may be attributed to their simulation logic, which emphasizes the number of nodes an influencer can reach and the extent of their subsequent network impact, without taking into account the semantic alignment between the influencer's content and the products.
Consequently, many influencers in the network may share the same interest domain but are still unsuitable for the product like Influencer 2 in the case study of Figure \ref{fig:show_case}.
The performance of GPT-4 w/ profile\&CoT varies across products due to differences in their social networks. 
Datasets like Spark Thinking feature large, information-rich networks with 490k interactions and a follower variance of over 130, enabling effective influencer identification. 
In contrast, the smaller interaction frequency and a low variance of 98 in the SUPOR Boosted Showerhead dataset lead to poorer performance, as the LLM struggles to extract meaningful insights and distinguish influencers.

\subsection{Analysis of LLM-based Methods}
In this section, we mainly discuss the performance of LLMs due to their state-of-the-art performance in sophisticated text processing.

\subsubsection{Case study}. In Figure~\ref{fig:show_case}, 
we compare the selected influencer's advertisement with the profiles and historical interactions of sampled influencees, along with their simulated behaviors and purchase likelihood.
For the chosen Influencer 1, the advertisement focuses on young child education, aligning well with its profile. 
Examining its Follower 1, we observe past interactions with the influencer on this product, and a history of strong support for the product. 
Consequently, this user will probably maintain their support, reflected in a predicted likelihood of 7. 
In contrast, the selected Influencer 2 has a mismatch as his followers are interested in pregnancy, not align with the ad that focuses on parenting.
Therefore, although the influencees have shown strong support for previous posts, GPT-4 predicts they will not support this product.
In Figure~\ref{fig:bar}(b), we present the normalized average distributions of comments and purchase likelihood for the top 10 influencers versus others. Top influencers attract more comments and higher purchase intentions.
Two illustrative examples highlight this contrast: Comments from influencers outside the top 10 tend to be off-topic and less indicative of purchase intent (e.g., `Is there a business cooperation opportunity?' and `Can this course be used for fitness?' regarding the Spark Thinking education product).
Conversely, sample comments from the top influencers, such as `Spark Thinking, my top choice!' and `A great product, useful, I want to buy', demonstrate a genuine interest in the product.
These cases demonstrate the role of text in simulating and predicting varying responses from influencees by considering multiple factors.

\subsubsection{Ablation of Human Preferences and Behavioral Patterns}
We conduct two extra modules to enhance LLM-based method. 
Firstly, compared with directly feeding influencer and influencee information into behavior prediction, we integrate the profile generation module in GPT-4 w/ Profile to provide the human preference of influencees, offering a personal profile of whether the influencees are interested in the posts the influencer makes and take corresponding actions.
Secondly, in GPT-4 with profile\&CoT, we incorporate an additional procedural introduction for behavior prediction, combining behavioral patterns with step-by-step generation prompts. Results in Table~\ref{tab:main} show each component's critical role in enhancing overall effectiveness.

\begin{figure}[htb]
    \centering
    \includegraphics[width=0.8\linewidth]{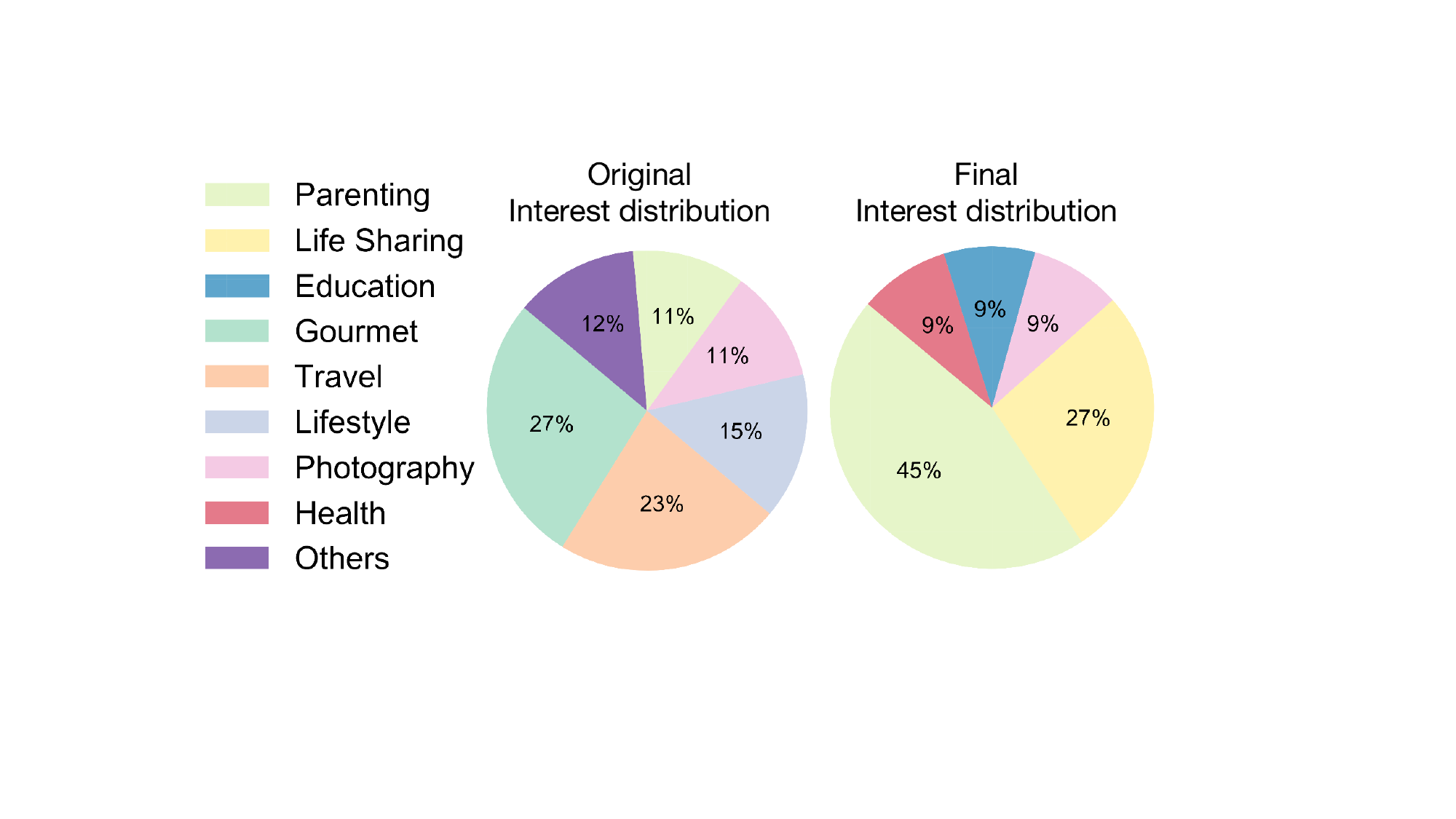}
    \caption{The interest distribution for the product `Spark Thinking' at the start and the final stage.
    The proportions of `parenting' and `life sharing' increase, consistent with the product's educational nature.
    }
    \label{fig:pie}
\end{figure}

\subsubsection{Interest develoment visualization}. In Figure~\ref{fig:pie}, we present the evolution of candidate influencer interests from their original state to the final distribution of GPT-4 w/ profile\&CoT for the product `Spark Thinking', an educational product for children. 
Initially, the seed influencers in our dataset span diverse areas like Gourmet and Travel, which are not directly related to our product. 
However, after the simulation, the percentage of influencers focused on parenting increased from 15\% to 40\%, life sharing from 20\% to 30\%, and education from 0\% to 10\%. 
This indicates that LLM's simulation effectively matches influencers from domains similar to those advertised for the product. 
Additionally, there are interests in other areas such as health and photography, suggesting that the matching process is not a simple keyword match within a single domain.

\begin{figure}[htb]
    \centering
    \includegraphics[width=0.8\linewidth]{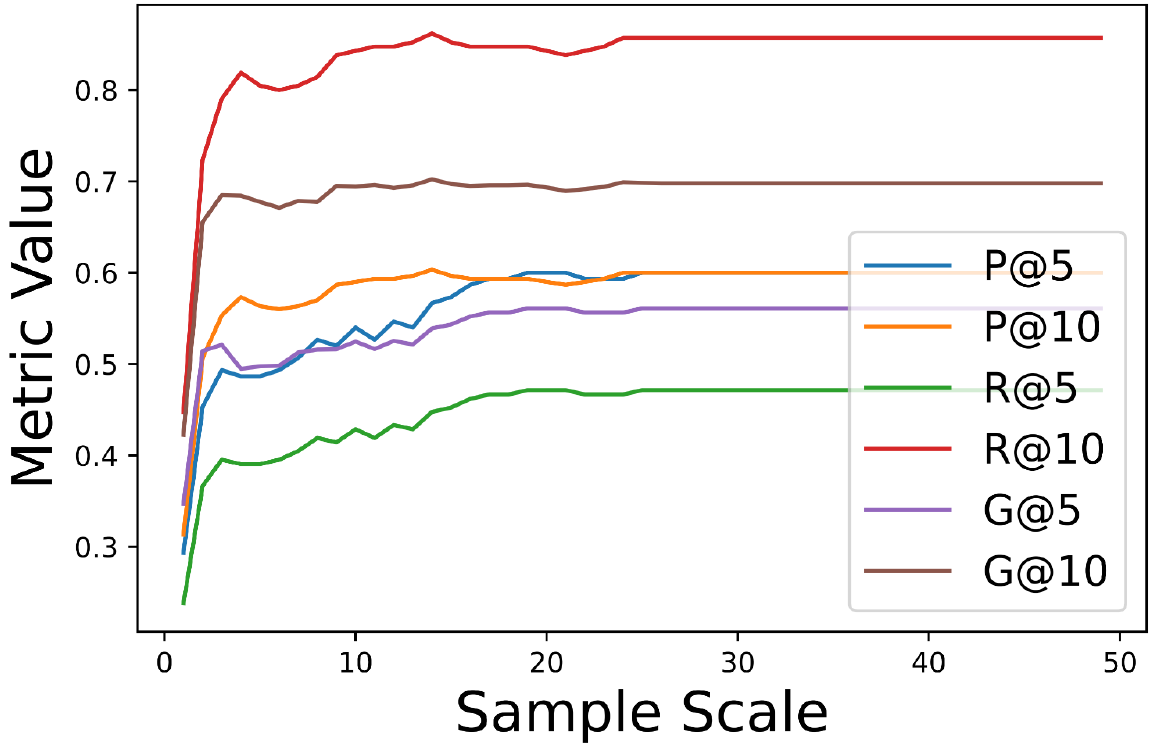}
    \caption{Performance on different simulation scales.
    }
    \label{fig:sample}
\end{figure}

\subsubsection{Simulation scale}. We conduct a parameter study to investigate the impact of varying simulation scales, specifically focusing on the number of sampled simulated influencees for each seed influencer. 
Illustrated in Figure~\ref{fig:sample}, 
we found that the performance trajectory initially demonstrates fluctuating growth, which then stabilizes upon exceeding a count of 20 influencees.  
These findings support our choice to use a sample size between 30 and 50, finding a good balance between capturing enough detail and keeping our simulations practical to run. This careful choice helps make sure our simulations are reliable and fit with our aim to accurately understand how influencers affect the digital world.

\begin{figure}[htb]
    \centering
    \includegraphics[width=0.7\linewidth]{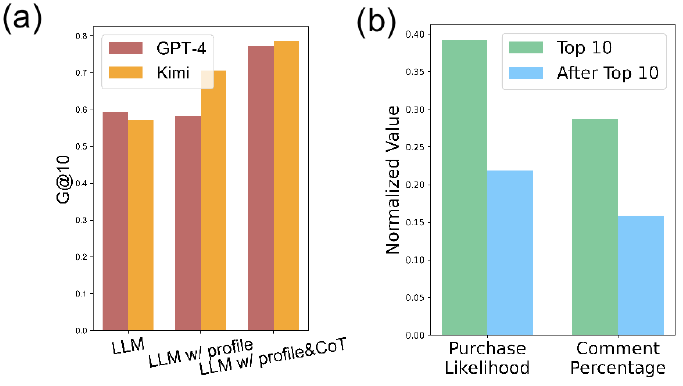}
    \caption{(a) Comparison of GPT-4 and Kimi results with profile\&CoT module.
    (b) The percentage of comments from influencees of the top 10 influencers versus those after the top 10, indicating higher engagement among the top 10's influencees.
    }
    \label{fig:bar}
\end{figure}

\subsubsection{Performance on different LLMs}. To showcase the broad applicability of our framework, which is not limited to specific LLM backbones, we experiment with alternative open APIs such as Kimi, developed by Moonshot AI. 
The results are presented in Figure~\ref{fig:bar}(a). 
It is evident that both models perform well, with Kimi exhibiting a slight edge in performance.
The comparison result further demonstrates the effectiveness of various components when working with Kimi.
This advantage could potentially be attributed to Kimi's tailored design for the Chinese language.

\section{Conclusion}
We introduce SAGraph, the first comprehensive, text-rich large-scale social advertisement graph dataset, designed to simulate influencer selection and optimize the effectiveness of advertising campaigns based on social networks.
It includes static social relationships and dynamic interactions towards specific posts, capturing key factors of influencer selection from digital associations to textual connotations, enabling large language models to be more naturally applied in advertising tracks, and creating more possibilities. 
Our experiment has demonstrated the effectiveness of LLM-based methods over traditional approaches in influencer selection.
We hope researchers will engage in the study of SAGraph, clarifying how to scientifically simulate the spread of influence for specific products on social networks, and opening up an exciting future for advertising campaigns in the era of prevailing self-media.

\bibliographystyle{ACM-Reference-Format}
\bibliography{sample-base}


\begin{thebibliography}{34}


\ifx \showCODEN    \undefined \def \showCODEN     #1{\unskip}     \fi
\ifx \showISBNx    \undefined \def \showISBNx     #1{\unskip}     \fi
\ifx \showISBNxiii \undefined \def \showISBNxiii  #1{\unskip}     \fi
\ifx \showISSN     \undefined \def \showISSN      #1{\unskip}     \fi
\ifx \showLCCN     \undefined \def \showLCCN      #1{\unskip}     \fi
\ifx \shownote     \undefined \def \shownote      #1{#1}          \fi
\ifx \showarticletitle \undefined \def \showarticletitle #1{#1}   \fi
\ifx \showURL      \undefined \def \showURL       {\relax}        \fi
\providecommand\bibfield[2]{#2}
\providecommand\bibinfo[2]{#2}
\providecommand\natexlab[1]{#1}
\providecommand\showeprint[2][]{arXiv:#2}

\bibitem[Borgs et~al\mbox{.}(2014)]%
        {borgs2014maximizing}
\bibfield{author}{\bibinfo{person}{Christian Borgs}, \bibinfo{person}{Michael Brautbar}, \bibinfo{person}{Jennifer Chayes}, {and} \bibinfo{person}{Brendan Lucier}.} \bibinfo{year}{2014}\natexlab{}.
\newblock \showarticletitle{Maximizing social influence in nearly optimal time}. In \bibinfo{booktitle}{\emph{Proceedings of the twenty-fifth annual ACM-SIAM symposium on Discrete algorithms}}. SIAM, \bibinfo{pages}{946--957}.
\newblock


\bibitem[Cao et~al\mbox{.}(2020)]%
        {cao2020popularity}
\bibfield{author}{\bibinfo{person}{Qi Cao}, \bibinfo{person}{Huawei Shen}, \bibinfo{person}{Jinhua Gao}, \bibinfo{person}{Bingzheng Wei}, {and} \bibinfo{person}{Xueqi Cheng}.} \bibinfo{year}{2020}\natexlab{}.
\newblock \showarticletitle{Popularity prediction on social platforms with coupled graph neural networks}. In \bibinfo{booktitle}{\emph{Proceedings of the 13th international conference on web search and data mining}}. \bibinfo{pages}{70--78}.
\newblock


\bibitem[{Covington \& Burling LLP}(2019)]%
        {datasecuritymanagement2019}
\bibfield{author}{\bibinfo{person}{{Covington \& Burling LLP}}.} \bibinfo{year}{2019}\natexlab{}.
\newblock \bibinfo{booktitle}{\emph{{Measures for Data Security Management}}}.
\newblock \bibinfo{type}{{T}echnical {R}eport} {Draft for Comments}. \bibinfo{institution}{{Covington \& Burling LLP}}, \bibinfo{address}{{Washington, D.C.}}
\newblock
\urldef\tempurl%
\url{{https://www.insideprivacy.com/wp-content/uploads/sites/51/2019/05/Measures-for-Data-Security-Management_Bilingual-1.pdf}}
\showURL{%
\tempurl}


\bibitem[Dutta et~al\mbox{.}(2020)]%
        {dutta2020deep}
\bibfield{author}{\bibinfo{person}{Subhabrata Dutta}, \bibinfo{person}{Sarah Masud}, \bibinfo{person}{Soumen Chakrabarti}, {and} \bibinfo{person}{Tanmoy Chakraborty}.} \bibinfo{year}{2020}\natexlab{}.
\newblock \showarticletitle{Deep exogenous and endogenous influence combination for social chatter intensity prediction}. In \bibinfo{booktitle}{\emph{Proceedings of the 26th ACM SIGKDD International Conference on Knowledge Discovery \& Data Mining}}. \bibinfo{pages}{1999--2008}.
\newblock


\bibitem[Galhotra et~al\mbox{.}(2016)]%
        {galhotra2016holistic}
\bibfield{author}{\bibinfo{person}{Sainyam Galhotra}, \bibinfo{person}{Akhil Arora}, {and} \bibinfo{person}{Shourya Roy}.} \bibinfo{year}{2016}\natexlab{}.
\newblock \showarticletitle{Holistic influence maximization: Combining scalability and efficiency with opinion-aware models}. In \bibinfo{booktitle}{\emph{Proceedings of the 2016 international conference on management of data}}. \bibinfo{pages}{743--758}.
\newblock


\bibitem[Gao et~al\mbox{.}(2019)]%
        {gao2019abstractive}
\bibfield{author}{\bibinfo{person}{Shen Gao}, \bibinfo{person}{Xiuying Chen}, \bibinfo{person}{Piji Li}, \bibinfo{person}{Zhaochun Ren}, \bibinfo{person}{Lidong Bing}, \bibinfo{person}{Dongyan Zhao}, {and} \bibinfo{person}{Rui Yan}.} \bibinfo{year}{2019}\natexlab{}.
\newblock \showarticletitle{Abstractive text summarization by incorporating reader comments}. In \bibinfo{booktitle}{\emph{Proceedings of the AAAI Conference on Artificial Intelligence}}, Vol.~\bibinfo{volume}{33}. \bibinfo{pages}{6399--6406}.
\newblock


\bibitem[Goyal et~al\mbox{.}(2011)]%
        {goyal2011celf++}
\bibfield{author}{\bibinfo{person}{Amit Goyal}, \bibinfo{person}{Wei Lu}, {and} \bibinfo{person}{Laks~VS Lakshmanan}.} \bibinfo{year}{2011}\natexlab{}.
\newblock \showarticletitle{Celf++ optimizing the greedy algorithm for influence maximization in social networks}. In \bibinfo{booktitle}{\emph{Proceedings of the 20th international conference companion on World wide web}}. \bibinfo{pages}{47--48}.
\newblock


\bibitem[Guare(2016)]%
        {guare2016six}
\bibfield{author}{\bibinfo{person}{John Guare}.} \bibinfo{year}{2016}\natexlab{}.
\newblock \showarticletitle{Six degrees of separation}.
\newblock In \bibinfo{booktitle}{\emph{The Contemporary Monologue: Men}}. \bibinfo{publisher}{Routledge}, \bibinfo{pages}{89--93}.
\newblock


\bibitem[Kempe et~al\mbox{.}(2003)]%
        {kempe2003maximizing}
\bibfield{author}{\bibinfo{person}{David Kempe}, \bibinfo{person}{Jon Kleinberg}, {and} \bibinfo{person}{{\'E}va Tardos}.} \bibinfo{year}{2003}\natexlab{}.
\newblock \showarticletitle{Maximizing the spread of influence through a social network}. In \bibinfo{booktitle}{\emph{Proceedings of the ninth ACM SIGKDD international conference on Knowledge discovery and data mining}}. \bibinfo{pages}{137--146}.
\newblock


\bibitem[Kwak(2010)]%
        {kwak2010twitter}
\bibfield{author}{\bibinfo{person}{H Kwak}.} \bibinfo{year}{2010}\natexlab{}.
\newblock \showarticletitle{What is Twitter, a social network or a news media?}
\newblock \bibinfo{journal}{\emph{Department of Computer Science, KAIST}} (\bibinfo{year}{2010}).
\newblock


\bibitem[Lahuerta-Otero and Cordero-Guti{\'e}rrez(2016)]%
        {lahuerta2016looking}
\bibfield{author}{\bibinfo{person}{Eva Lahuerta-Otero} {and} \bibinfo{person}{Rebeca Cordero-Guti{\'e}rrez}.} \bibinfo{year}{2016}\natexlab{}.
\newblock \showarticletitle{Looking for the perfect tweet. The use of data mining techniques to find influencers on twitter}.
\newblock \bibinfo{journal}{\emph{Computers in human behavior}}  \bibinfo{volume}{64} (\bibinfo{year}{2016}), \bibinfo{pages}{575--583}.
\newblock


\bibitem[Lampos et~al\mbox{.}(2014)]%
        {lampos2014predicting}
\bibfield{author}{\bibinfo{person}{Vasileios Lampos}, \bibinfo{person}{Nikolaos Aletras}, \bibinfo{person}{Daniel Preo{\c{t}}iuc-Pietro}, {and} \bibinfo{person}{Trevor Cohn}.} \bibinfo{year}{2014}\natexlab{}.
\newblock \showarticletitle{Predicting and characterising user impact on Twitter}. In \bibinfo{booktitle}{\emph{Proceedings of the 14th Conference of the European Chapter of the Association for Computational Linguistics}}. \bibinfo{pages}{405--413}.
\newblock


\bibitem[Lenger(2022)]%
        {lenger2022choose}
\bibfield{author}{\bibinfo{person}{Asl{\i}~Diyadin Lenger}.} \bibinfo{year}{2022}\natexlab{}.
\newblock \showarticletitle{How to choose the right influencer for a marketing strategy}.
\newblock \bibinfo{journal}{\emph{Applied Marketing Analytics}} \bibinfo{volume}{8}, \bibinfo{number}{1} (\bibinfo{year}{2022}), \bibinfo{pages}{89--104}.
\newblock


\bibitem[Leskovec et~al\mbox{.}(2007a)]%
        {leskovec2007graph}
\bibfield{author}{\bibinfo{person}{Jure Leskovec}, \bibinfo{person}{Jon Kleinberg}, {and} \bibinfo{person}{Christos Faloutsos}.} \bibinfo{year}{2007}\natexlab{a}.
\newblock \showarticletitle{Graph evolution: Densification and shrinking diameters}.
\newblock \bibinfo{journal}{\emph{ACM transactions on Knowledge Discovery from Data (TKDD)}} \bibinfo{volume}{1}, \bibinfo{number}{1} (\bibinfo{year}{2007}), \bibinfo{pages}{2--es}.
\newblock


\bibitem[Leskovec et~al\mbox{.}(2007b)]%
        {leskovec2007cost}
\bibfield{author}{\bibinfo{person}{Jure Leskovec}, \bibinfo{person}{Andreas Krause}, \bibinfo{person}{Carlos Guestrin}, \bibinfo{person}{Christos Faloutsos}, \bibinfo{person}{Jeanne VanBriesen}, {and} \bibinfo{person}{Natalie Glance}.} \bibinfo{year}{2007}\natexlab{b}.
\newblock \showarticletitle{Cost-effective outbreak detection in networks}. In \bibinfo{booktitle}{\emph{Proceedings of the 13th ACM SIGKDD international conference on Knowledge discovery and data mining}}. \bibinfo{pages}{420--429}.
\newblock


\bibitem[Leskovec and Mcauley(2012)]%
        {leskovec2012learning}
\bibfield{author}{\bibinfo{person}{Jure Leskovec} {and} \bibinfo{person}{Julian Mcauley}.} \bibinfo{year}{2012}\natexlab{}.
\newblock \showarticletitle{Learning to discover social circles in ego networks}.
\newblock \bibinfo{journal}{\emph{Advances in neural information processing systems}}  \bibinfo{volume}{25} (\bibinfo{year}{2012}).
\newblock


\bibitem[Leskovec and Sosic(2014)]%
        {leskovec2014snap}
\bibfield{author}{\bibinfo{person}{Jure Leskovec} {and} \bibinfo{person}{Rok Sosic}.} \bibinfo{year}{2014}\natexlab{}.
\newblock \bibinfo{title}{SNAP: A general purpose network analysis and graph mining library in C++}.
\newblock


\bibitem[Leskovec and Sosi{\v{c}}(2016)]%
        {leskovec2016snap}
\bibfield{author}{\bibinfo{person}{Jure Leskovec} {and} \bibinfo{person}{Rok Sosi{\v{c}}}.} \bibinfo{year}{2016}\natexlab{}.
\newblock \showarticletitle{Snap: A general-purpose network analysis and graph-mining library}.
\newblock \bibinfo{journal}{\emph{ACM Transactions on Intelligent Systems and Technology (TIST)}} \bibinfo{volume}{8}, \bibinfo{number}{1} (\bibinfo{year}{2016}), \bibinfo{pages}{1--20}.
\newblock


\bibitem[Li et~al\mbox{.}(2020b)]%
        {li2020data}
\bibfield{author}{\bibinfo{person}{Jiawei Li}, \bibinfo{person}{Qing Xu}, \bibinfo{person}{Raphael Cuomo}, \bibinfo{person}{Vidya Purushothaman}, \bibinfo{person}{Tim Mackey}, {et~al\mbox{.}}} \bibinfo{year}{2020}\natexlab{b}.
\newblock \showarticletitle{Data mining and content analysis of the Chinese social media platform Weibo during the early COVID-19 outbreak: retrospective observational infoveillance study}.
\newblock \bibinfo{journal}{\emph{JMIR Public Health and Surveillance}} \bibinfo{volume}{6}, \bibinfo{number}{2} (\bibinfo{year}{2020}), \bibinfo{pages}{e18700}.
\newblock


\bibitem[Li et~al\mbox{.}(2020a)]%
        {li2020vmsmo}
\bibfield{author}{\bibinfo{person}{Mingzhe Li}, \bibinfo{person}{Xiuying Chen}, \bibinfo{person}{Shen Gao}, \bibinfo{person}{Zhangming Chan}, \bibinfo{person}{Dongyan Zhao}, {and} \bibinfo{person}{Rui Yan}.} \bibinfo{year}{2020}\natexlab{a}.
\newblock \showarticletitle{VMSMO: Learning to Generate Multimodal Summary for Video-based News Articles}. In \bibinfo{booktitle}{\emph{Proceedings of the 2020 Conference on Empirical Methods in Natural Language Processing (EMNLP)}}. \bibinfo{pages}{9360--9369}.
\newblock


\bibitem[Li et~al\mbox{.}(2018)]%
        {li2018retweeting}
\bibfield{author}{\bibinfo{person}{Qian Li}, \bibinfo{person}{Xiaojuan Li}, \bibinfo{person}{Bin Wu}, {and} \bibinfo{person}{Yunpeng Xiao}.} \bibinfo{year}{2018}\natexlab{}.
\newblock \showarticletitle{Retweeting prediction based on social hotspots and dynamic tensor decomposition}.
\newblock \bibinfo{journal}{\emph{IEICE TRANSACTIONS on Information and Systems}} \bibinfo{volume}{101}, \bibinfo{number}{5} (\bibinfo{year}{2018}), \bibinfo{pages}{1380--1392}.
\newblock


\bibitem[Li et~al\mbox{.}(2015)]%
        {li2015real}
\bibfield{author}{\bibinfo{person}{Yuchen Li}, \bibinfo{person}{Dongxiang Zhang}, {and} \bibinfo{person}{Kian-Lee Tan}.} \bibinfo{year}{2015}\natexlab{}.
\newblock \showarticletitle{Real-time targeted influence maximization for online advertisements}.
\newblock  (\bibinfo{year}{2015}).
\newblock


\bibitem[Mallipeddi et~al\mbox{.}(2022)]%
        {mallipeddi2022framework}
\bibfield{author}{\bibinfo{person}{Rakesh~R Mallipeddi}, \bibinfo{person}{Subodha Kumar}, \bibinfo{person}{Chelliah Sriskandarajah}, {and} \bibinfo{person}{Yunxia Zhu}.} \bibinfo{year}{2022}\natexlab{}.
\newblock \showarticletitle{A framework for analyzing influencer marketing in social networks: selection and scheduling of influencers}.
\newblock \bibinfo{journal}{\emph{Management Science}} \bibinfo{volume}{68}, \bibinfo{number}{1} (\bibinfo{year}{2022}), \bibinfo{pages}{75--104}.
\newblock


\bibitem[Manlio De~Domenico and Latora(2013)]%
        {Centrality}
\bibfield{author}{\bibinfo{person}{Alex~Arenas Manlio De~Domenico, Vincenzo~Nicosia} {and} \bibinfo{person}{Vito Latora}.} \bibinfo{year}{2013}\natexlab{}.
\newblock \showarticletitle{Centrality measures in multilayer networks}.
\newblock  (\bibinfo{year}{2013}).
\newblock


\bibitem[Qiu et~al\mbox{.}(2018)]%
        {qiu2018deepinf}
\bibfield{author}{\bibinfo{person}{Jiezhong Qiu}, \bibinfo{person}{Jian Tang}, \bibinfo{person}{Hao Ma}, \bibinfo{person}{Yuxiao Dong}, \bibinfo{person}{Kuansan Wang}, {and} \bibinfo{person}{Jie Tang}.} \bibinfo{year}{2018}\natexlab{}.
\newblock \showarticletitle{Deepinf: Social influence prediction with deep learning}. In \bibinfo{booktitle}{\emph{Proceedings of the 24th ACM SIGKDD international conference on knowledge discovery \& data mining}}. \bibinfo{pages}{2110--2119}.
\newblock


\bibitem[Rivadeneira et~al\mbox{.}(2021)]%
        {rivadeneira2021predicting}
\bibfield{author}{\bibinfo{person}{Luc{\'\i}a Rivadeneira}, \bibinfo{person}{Jian-Bo Yang}, {and} \bibinfo{person}{Manuel L{\'o}pez-Ib{\'a}{\~n}ez}.} \bibinfo{year}{2021}\natexlab{}.
\newblock \showarticletitle{Predicting tweet impact using a novel evidential reasoning prediction method}.
\newblock \bibinfo{journal}{\emph{Expert Systems with Applications}}  \bibinfo{volume}{169} (\bibinfo{year}{2021}), \bibinfo{pages}{114400}.
\newblock


\bibitem[Tang et~al\mbox{.}(2015)]%
        {tang2015influence}
\bibfield{author}{\bibinfo{person}{Youze Tang}, \bibinfo{person}{Yanchen Shi}, {and} \bibinfo{person}{Xiaokui Xiao}.} \bibinfo{year}{2015}\natexlab{}.
\newblock \showarticletitle{Influence maximization in near-linear time: A martingale approach}. In \bibinfo{booktitle}{\emph{Proceedings of the 2015 ACM SIGMOD international conference on management of data}}. \bibinfo{pages}{1539--1554}.
\newblock


\bibitem[Tang et~al\mbox{.}(2014)]%
        {tang2014influence}
\bibfield{author}{\bibinfo{person}{Youze Tang}, \bibinfo{person}{Xiaokui Xiao}, {and} \bibinfo{person}{Yanchen Shi}.} \bibinfo{year}{2014}\natexlab{}.
\newblock \showarticletitle{Influence maximization: Near-optimal time complexity meets practical efficiency}. In \bibinfo{booktitle}{\emph{Proceedings of the 2014 ACM SIGMOD international conference on Management of data}}. \bibinfo{pages}{75--86}.
\newblock


\bibitem[Yan et~al\mbox{.}(2019)]%
        {yan2019minimizing}
\bibfield{author}{\bibinfo{person}{Ruidong Yan}, \bibinfo{person}{Deying Li}, \bibinfo{person}{Weili Wu}, \bibinfo{person}{Ding-Zhu Du}, {and} \bibinfo{person}{Yongcai Wang}.} \bibinfo{year}{2019}\natexlab{}.
\newblock \showarticletitle{Minimizing influence of rumors by blockers on social networks: algorithms and analysis}.
\newblock \bibinfo{journal}{\emph{IEEE Transactions on Network Science and Engineering}} \bibinfo{volume}{7}, \bibinfo{number}{3} (\bibinfo{year}{2019}), \bibinfo{pages}{1067--1078}.
\newblock


\bibitem[Yang et~al\mbox{.}(2012)]%
        {yang2012automatic}
\bibfield{author}{\bibinfo{person}{Fan Yang}, \bibinfo{person}{Yang Liu}, \bibinfo{person}{Xiaohui Yu}, {and} \bibinfo{person}{Min Yang}.} \bibinfo{year}{2012}\natexlab{}.
\newblock \showarticletitle{Automatic detection of rumor on sina weibo}. In \bibinfo{booktitle}{\emph{Proceedings of the ACM SIGKDD workshop on mining data semantics}}. \bibinfo{pages}{1--7}.
\newblock


\bibitem[Zhang et~al\mbox{.}(2013)]%
        {zhang2013social}
\bibfield{author}{\bibinfo{person}{Jing Zhang}, \bibinfo{person}{Biao Liu}, \bibinfo{person}{Jie Tang}, \bibinfo{person}{Ting Chen}, {and} \bibinfo{person}{Juanzi Li}.} \bibinfo{year}{2013}\natexlab{}.
\newblock \showarticletitle{Social influence locality for modeling retweeting behaviors}. In \bibinfo{booktitle}{\emph{Twenty-third international joint conference on artificial intelligence}}. Citeseer.
\newblock


\bibitem[Zhang et~al\mbox{.}(2015)]%
        {zhang2015influenced}
\bibfield{author}{\bibinfo{person}{Jing Zhang}, \bibinfo{person}{Jie Tang}, \bibinfo{person}{Juanzi Li}, \bibinfo{person}{Yang Liu}, {and} \bibinfo{person}{Chunxiao Xing}.} \bibinfo{year}{2015}\natexlab{}.
\newblock \showarticletitle{Who influenced you? predicting retweet via social influence locality}.
\newblock \bibinfo{journal}{\emph{ACM Transactions on Knowledge Discovery from Data (TKDD)}} \bibinfo{volume}{9}, \bibinfo{number}{3} (\bibinfo{year}{2015}), \bibinfo{pages}{1--26}.
\newblock


\bibitem[Zhang et~al\mbox{.}(2022)]%
        {zhang2022blocking}
\bibfield{author}{\bibinfo{person}{Zonghan Zhang}, \bibinfo{person}{Subhodip Biswas}, \bibinfo{person}{Fanglan Chen}, \bibinfo{person}{Kaiqun Fu}, \bibinfo{person}{Taoran Ji}, \bibinfo{person}{Chang-Tien Lu}, \bibinfo{person}{Naren Ramakrishnan}, {and} \bibinfo{person}{Zhiqian Chen}.} \bibinfo{year}{2022}\natexlab{}.
\newblock \showarticletitle{Blocking Influence at Collective Level with Hard Constraints (Student Abstract)}. In \bibinfo{booktitle}{\emph{Proceedings of the AAAI Conference on Artificial Intelligence}}, Vol.~\bibinfo{volume}{36}. \bibinfo{pages}{13115--13116}.
\newblock


\bibitem[Zhong et~al\mbox{.}(2018)]%
        {zhong2018identifying}
\bibfield{author}{\bibinfo{person}{Lin-Feng Zhong}, \bibinfo{person}{Ming-Sheng Shang}, \bibinfo{person}{Xiao-Long Chen}, {and} \bibinfo{person}{Shi-Ming Cai}.} \bibinfo{year}{2018}\natexlab{}.
\newblock \showarticletitle{Identifying the influential nodes via eigen-centrality from the differences and similarities of structure}.
\newblock \bibinfo{journal}{\emph{Physica A: Statistical Mechanics and its Applications}}  \bibinfo{volume}{510} (\bibinfo{year}{2018}), \bibinfo{pages}{77--82}.
\newblock


\end{thebibliography}


\end{document}